\documentclass[12pt, authoryear]{elsarticle}

\makeatletter
\def\ps@pprintTitle{%
 \let\@oddhead\@empty
 \let\@evenhead\@empty
 \def\@oddfoot{\centerline{\thepage}}%
 \let\@evenfoot\@oddfoot}
\makeatother

\usepackage{amsmath}
\usepackage{graphicx}
\usepackage{verbatim}
\usepackage{subfigure}
\usepackage{color}
\usepackage{setspace}
\usepackage{float}
\usepackage{natbib}
\usepackage{footnote}
\usepackage{lineno}
\usepackage{url}
\makesavenoteenv{tabular}
\makesavenoteenv{table}

\usepackage[top=2.3cm,left=2.3cm,right=2.3cm,bottom=2.3cm]{geometry}

\begin{document}

\begin{frontmatter}

\title{Measuring temperature and ammonia hydrate ice on Charon in 2015 from
  Keck/OSIRIS spectra}

\author[lasp]{B. J. Holler}
\author[swri]{L. A. Young}
\author[swri]{M. W. Buie}
\author[low]{W. M. Grundy}
\author[keck]{J. E. Lyke}
\author[swri]{E. F. Young}
\author[low]{H. G. Roe}

\address[lasp]{Laboratory for Atmospheric and Space Physics, University of Colorado at Boulder, 1234 Innovation Dr., Boulder, CO 80303.}
\address[swri]{Southwest Research Institute, 1050 Walnut St. \#300,
Boulder, CO 80302.}
\address[low]{Lowell Observatory, 1400 W. Mars Hill Rd., Flagstaff, AZ
86001.}
\address[keck]{W. M. Keck Observatory, 65-1120 Mamalahoa Hwy.,
Kamuela, HI 96743.}

\begin{abstract}
\onehalfspacing{In this work we investigated the longitudinal (zonal)
  variability of H$_2$O and ammonia (NH$_3$) hydrate ices on the surface of
  Charon through analysis of the 1.65 $\mu$m and 2.21 $\mu$m
  absorption features, respectively. Near-infrared spectra  presented here were obtained
between 2015-07-14 and 2015-08-30 UT with the OSIRIS integral field
spectrograph on Keck I. Spectra centered on six different sub-observer longitudes were
obtained through the Hbb (1.473-1.803 $\mu$m) and Kbb (1.965-2.381
$\mu$m) filters. Gaussian functions were
fit to the aforementioned bands to obtain information on band center,
band depth, full width at half maximum, and band area. The shift in the
band center of the temperature-dependent 1.65 $\mu$m feature was used
to calculate the H$_2$O ice temperature. The mean temperature of the
ice on the observable portion of Charon's surface is 45$\pm$14 K and we
report no statistically significant variations in temperature across
the surface. We hypothesize that the crystalline
and amorphous phases of water ice reached equilibrium over 3.5 Gyr
ago, with thermal recrystallization balancing the effects of
irradiation amorphization. We do not believe that cryovolcanism is
necessary to explain the presence of crystalline water ice on the
surface of Charon. Absorption from ammonia species is detected between
12$^{\circ}$ and 290$^{\circ}$, in agreement with results from New
Horizons. Ongoing diffusion of ammonia through the rocky
mantle and upper layer of water ice is one possible mechanism for
maintaining its presence in Charon's surface ice. Reduced Charon spectra
corrected for telluric and solar absorption are available as
supplementary online material.}
\end{abstract}


\end{frontmatter}

\section{Introduction}
\onehalfspace{Charon, the largest moon of Pluto, was
serendipitously discovered in 1978 as an unresolved extension to Pluto's disk
that predictably changed position (Christy and Harrington,
1978). In the nearly four decades since its discovery, and especially since
the mutual events in the late 1980s (Buie et al., 1987; Marcialis et
al., 1987), studies focused solely on Charon remain sparse due largely
to the difficulty in isolating Charon from Pluto from most
ground-based observatories. Charon
orbits the system barycenter at a distance of
19,750 km, or $\sim$17 R$_{Pluto}$ (Tholen et al., 2008; Brozovi\'c et
al., 2015); this translates to a maximum angular separation of less
than 1$''$ at Pluto and Charon's geocentric distance in 2015 ($\sim$32
AU). Spectral observations of Charon were
once best suited for space-based facilities, however, ground-based facilities with large-class
telescopes and adaptive optics systems can now effectively obtain
unblended spectra of Pluto and Charon at these small
separations. Observations of Charon from the New Horizons flyby
provide spatially resolved ``ground truth'' context for previous and future Earth-based observations.\\
\indent The near-infrared spectrum of Charon is dominated by
absorption features of water ice (Brown and Calvin, 2000; Buie and Grundy, 2000; Dumas
et al., 2001), with over 90\% thought to be in the
crystalline phase (Cook et al., 2007; Merlin et al., 2010). The only
other confirmed ice species detected on
Charon is ammonia (NH$_3$) hydrate, with an absorption
feature at 2.21 $\mu$m (Brown and Calvin, 2000; Buie and Grundy, 2000;
Dumas et al., 2001; Cook et al., 2007; Verbiscer et al., 2007; Merlin
et al., 2010). Both crystalline water ice (Buie and Grundy, 2000;
Dumas et al., 2001) and ammonia hydrate (DeMeo et al., 2015) may show
evidence for differences in absorption between the leading and
trailing hemispheres. A difference in surface ice temperature of over 10 K
between the sub-Pluto and anti-Pluto hemispheres was calculated by Cook et
al. (2007); however, they assigned uncertainties of 5-10 K to the temperature
measurements, so the calculated difference may not be real.\\
\indent The lack of volatile ices (N$_2$, CH$_4$, and CO) on Charon is
consistent with theories of volatile loss and retention on small outer solar system
objects (Schaller and Brown, 2007; Johnson et al., 2015). At 45 K
(this work), the partial pressures of N$_2$ ($\sim$0.6 mbar), CH$_4$ ($\sim$0.2 $\mu$bar),
and CO ($\sim$0.1 mbar) are high enough to form an atmosphere (Fray and
Schmitt, 2009). New Horizons observations of a
solar occultation in the UV place 3-$\sigma$ upper limits
on Charon's surface pressure at 4.2, 0.3, and 1.2 picobar for
atmospheres composed solely of N$_2$, CH$_4$, and
CO, respectively (Stern et al., submitted). This means Charon's volatiles
escaped from its atmosphere and surface over the age of the solar
system; small quantities may exist on the surface, but if so,
are well below the threshold for detectability. In contrast, the
non-volatile ices H$_2$O and NH$_3$ are detected on Charon; due to their
negligible partial pressures (Fray and Schmitt, 2009), these ices are retained.\\
\indent Cook et al. (2007) define the $e$-folding time for the
conversion of crystalline water ice to amorphous water ice as the time
necessary to reduce the crystalline fraction (ratio of crystalline
water ice to total water ice) from 1 to 1/$e$. The crystalline
fraction is an exponential function and decreases with increasing
radiation dosage, which is a function of heliocentric distance and
time (Cooper et al., 2003). The $e$-folding time
for crystalline water ice to be converted to amorphous
water ice in a radiation environment of 1 eV-10 GeV protons,
without considering recrystallization processes, is 1.5 Myr; this is down to the depth
probed by near-infrared $H$ and $K$ band observations ($\sim$350
$\mu$m; Cook et al., 2007). The destruction of ammonia hydrate down to
this same depth in the same radiation environment is 20 Myr (Cooper et al., 2003; Cook et al.,
2007). The amorphous phase, which only makes up
$\sim$10\% of the water ice content on Charon (Cook et al., 2007;
Merlin et al., 2010), should dominate in high-radiation environments
since crystalline water ice requires a formation temperature greater
than 140 K (Leto and Baratta, 2003), well above the surface
temperature of Charon. However, laboratory work suggests that
thermal recrystallization achieves an equilibrium with irradiation
amorphization after a period of time that depends on
temperature (Leto and Baratta,
2003; Mastrapa and Brown, 2006; Zheng et al., 2009). The observed
fractions of crystalline and amorphous water
ice on Charon may be real and potentially explainable without invoking
a replenishment mechanism (see Results \& discussion).\\
\indent Crystalline water ice is formed in an orderly lattice structure while
amorphous water ice is a random organization of water molecules. This
structural difference results in distinctive absorption features between the
spectra of the two phases. In particular, an
absorption feature at 1.65 $\mu$m is present only in the spectrum of
crystalline water ice and is absent in the spectrum of amorphous water ice (Grundy
and Schmitt, 1998). This absorption feature is of special interest
because it is temperature-dependent: The shift of the band with
respect to a reference value provides a means of measuring the
temperature and was used in previous work to calculate the surface
temperature of Charon (e.g., Cook et al., 2007). Verbiscer et
al. (2006) modeled spectra of Charon and found that ammonia hydrate suppresses
the 1.65 $\mu$m crystalline H$_2$O band. It is possible that the
presence of ammonia hydrate in an intimate mixture with water ice could shift the
1.65 $\mu$m band center, but optical constants for such a mixture are
limited to laboratory measurements at 77 K (Brown et al., 1988). Lacking definite information, we
performed this work under the assumption that ammonia
hydrate does not shift the 1.65 $\mu$m band center.}

\section{Observations}
\onehalfspace{Pluto and Charon were observed on 6 nights in
Summer 2015 using the OH-Suppressing Infra-Red Imaging Spectrograph
(OSIRIS) on Keck I (Larkin et al., 2006; Mieda et al.,
2014). The motivation and design of these observations was to obtain
marginally resolved spectral cubes of Pluto at spectral resolutions 10
times higher than that from New Horizons. We present an analysis of
Charon here. Pluto will be addressed in a later paper. Observational
circumstances for each night are found in
Table 1 and the hemisphere of Charon visible on each night is shown
in Fig. 1. Over the 6 nights, a total of 240 minutes of $H$ band
observations and 260 minutes of $K$ band observations were
obtained. Spectra of Pluto and Charon were also obtained in 2010, 2012,
and 2013 but were not included in this work because we wished to focus
on those data taken during the New Horizons flyby epoch for direct
comparison to New Horizons results. This direct comparison places
ground-based observations in the context of the ``ground truth''
provided by New Horizons and will enhance the results of later
analysis of previous years' data.

\subsection{The OSIRIS instrument}
The OSIRIS instrument is a spectrograph equipped with an integral field unit (IFU), so it
obtains spectra at multiple spatial locations within the field of
view. We used OSIRIS in conjunction with the adaptive optics (AO)
system on Keck I to reduce the size of the point spread function (PSF) for Pluto and
Charon. The AO system was operated in laser guide star (LGS) mode on
one night (2015-07-14 UT) and in natural guide star (NGS) mode for the
other five nights. NGS mode was preferred and used more frequently
than LGS mode due to above-average seeing ($\sim$0.3$''$ in $H$) and because LGS
required a 45-60 minute laser check-out period. After passing through the AO
system, light from a rectangular region of sky passes through an array
of lenslets and onto a grating, producing overlapping spectra on the
detector. Each lenslet roughly corresponds to one pixel on the detector.

\begin{figure}[h!]
\begin{center}
\includegraphics[scale=0.85,trim=1.5cm 7cm 1cm 1.5cm,clip=true]{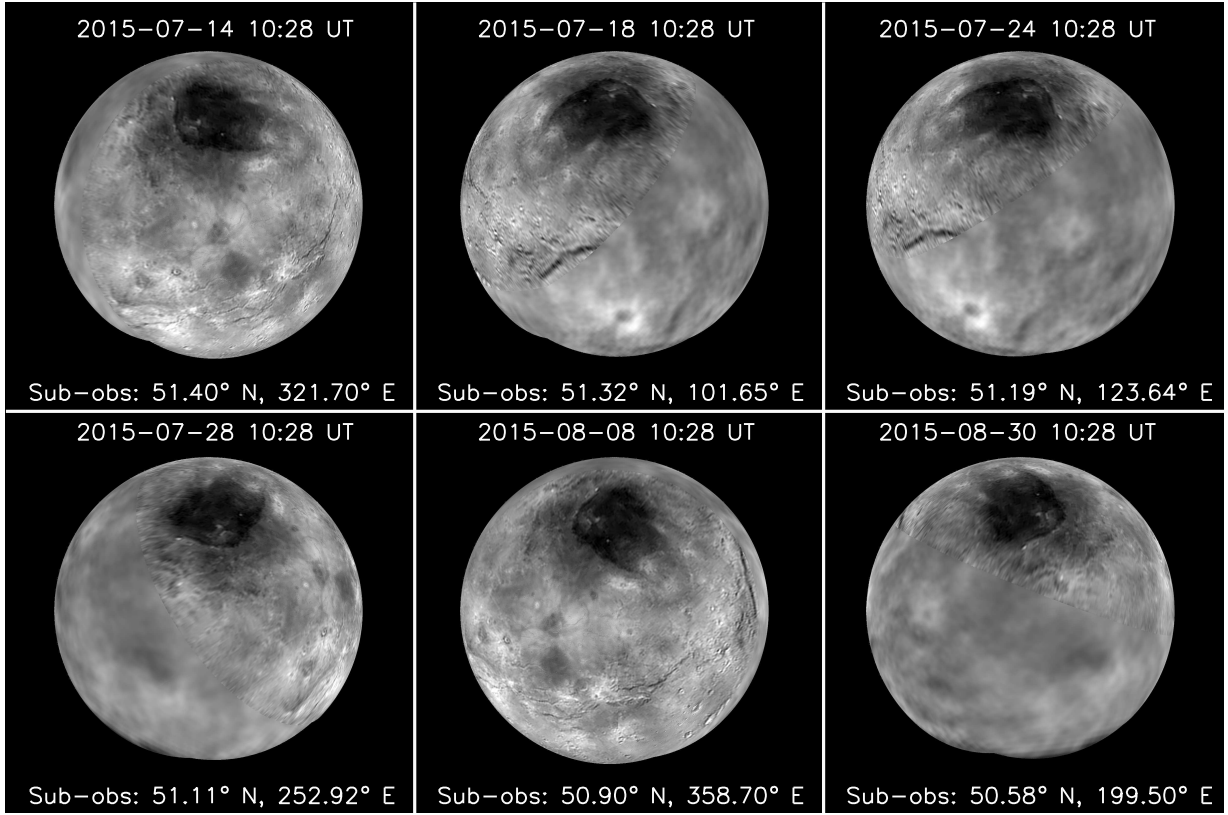}
\caption{Spherical projections of Charon showing the hemisphere visible from Earth at the UT date and
  time specified above each image. The sub-observer latitude and
  longitude, specified beneath each image, marks the point directly
  ``below'' the observer and therefore the center of each map. These spherical maps were
  constructed from a visible light Charon albedo map obtained by the New
  Horizons spacecraft and available from the JPL
  Photojournal (PIA19866: Global Map of Pluto's Moon Charon; http://photojournal.jpl.nasa.gov/catalog/PIA19866).}
\end{center}
\end{figure}

\indent We observed Pluto and Charon in the near-infrared through the
broadband Hbb ($H$ band, 1.473-1.803 $\mu$m) and Kbb ($K$ band,
1.965-2.381 $\mu$m) filters that use a 16$\times$64 lenslet array, for
a total of 1,019 overlapping spectra (five spectra fall off the edge
of the detector). The dispersion of the Hbb and Kbb filters is 0.000200
and 0.000250 $\mu$m/pixel, respectively, and the average spectral
resolution is $\sim$3800. For comparison, the spectral resolution of
the Linear Etalon Imaging Spectral Array (LEISA) on the New
Horizons spacecraft is 250 between 1.25 and 2.50 $\mu$m, with a
special region of higher resolution ($\lambda/\Delta\lambda$$\sim$560)
between 2.10 and 2.25 $\mu$m (Reuter et al., 2008; Young et al., 2008). OSIRIS therefore
provides a spectral resolving power over 15$\times$ higher than LEISA
over the full spectral range, and almost 7$\times$ higher over LEISA's
special region.}

\subsection{Observing strategy}
\onehalfspacing{Our observations were made using the finest
plate scale of 0.020$''$ per lenslet, resulting in a
0.32$''\times$1.28$''$ field of view. This field of view was aligned
along the imaginary line connecting Pluto and Charon, with Charon
always at the top of the field and Pluto always at the bottom (Fig. 2);
consistent alignment helped simplify the reduction process. In Summer
2015, Pluto and Charon were $\sim$32 AU from the Earth,
resulting in a spatial resolution of $\sim$500 km/pixel. At this
resolution, Charon was over two pixels in diameter and the centers of
the two bodies were separated by $>$30 pixels on the chip
(Fig. 2). Adaptive optics and the large
separation between Pluto and Charon significantly reduced cross-contamination so
separate spectra of each object could be extracted. The range of
angular separations during the observing period for Styx
(1.47-1.82$''$), Nix (1.60-2.19$''$), Kerberos (1.98-2.54$''$), and
Hydra (2.21-2.85$''$) placed them outside the field of view;
additionally, they are too faint to be detected by OSIRIS.\\
\indent An A-B-A-sky dither pattern was used to ensure that Pluto and
Charon did not fall on the same pixels throughout the night. The magnitude of the dither was small
enough that both Pluto and Charon were in the field of view in every image, even at
larger angular separations. The sky image from each set was used in the reduction to
remove sky background; AB subtraction was not performed, hence the use of
a small dither. The integration time for each component of the
dither sequence was 8 min in the Hbb filter and 5 min in the Kbb filter. The
integration time for the Kbb filter was shorter due to higher levels
of sky background in the $K$ band.\\
\indent These observations provided near-complete longitudinal
coverage of Charon over a 1.5 month span (Table 1). Six sets of
observations separated in sub-observer longitude by 60$^{\circ}$
provide ideal coverage, but this is difficult to perform in a short
time frame due to the 6.4-day rotation of Charon. Due to an
unanticipated failure of an important component of the OSIRIS
instrument in late June 2015, we were unable to observe on our preferred (and originally
scheduled) dates in early July. However, because it was crucial that these observations
be made around the time of the New Horizons flyby through the Pluto system,
our lost nights were rescheduled. This resulted in some regions of Charon being oversampled while
others were undersampled. The longitudinal coverage was still adequate
for the purposes of quantifying variability across the surface of
Charon. A solar analog was observed at the beginning of each night in both the Hbb and Kbb
filters to correct the Charon spectra for solar absorption
lines. Spectra of the solar analog HD 159662 (G2/G3 V spectral type;
Houk and Smith-Moore, 1988) were used for this purpose.}

\begin{figure}[h!]
\begin{center}
\includegraphics[scale=0.7,trim=0cm 5.85cm 0cm 2.2cm,clip=true]{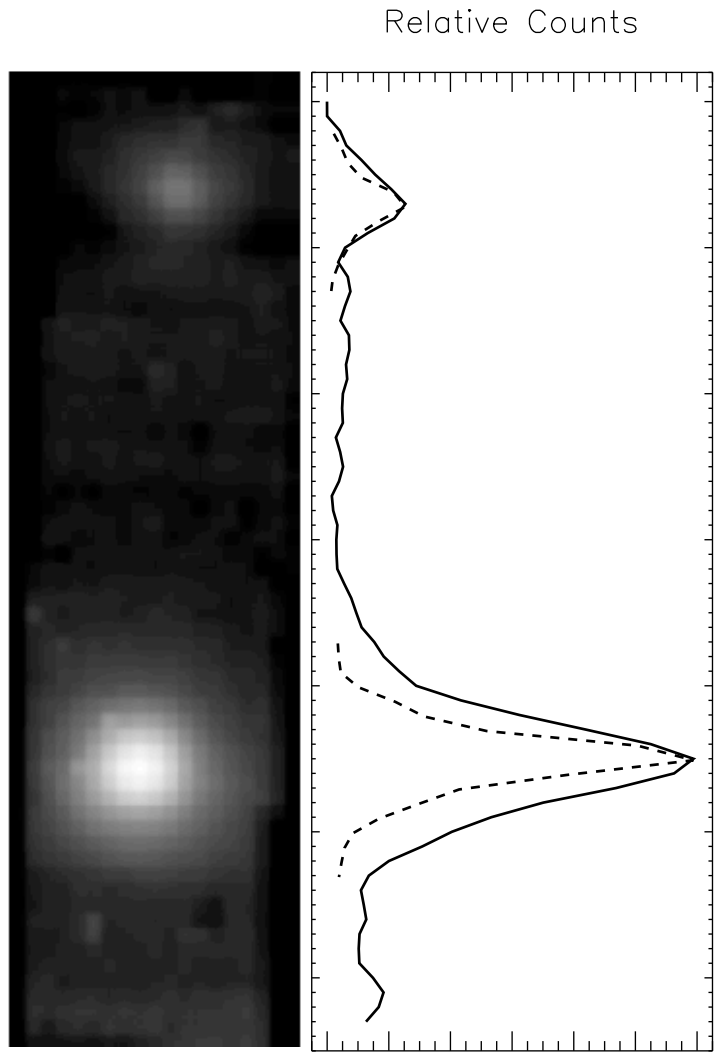}
\caption{\textit{Left:} Spectrally-averaged image of a 3D data cube obtained on the
  night of 2015-07-14 UT with the Kbb filter (1.965-2.381
  $\mu$m). These data were taken in the $B$ dither position with an
  exposure time of 300 seconds at an airmass of 1.50. Pluto is at the
  bottom of the image and Charon is near the top. The two
  are clearly separated on the detector even at an angular separation of
  0.74$''$. This image is representative of the data obtained
  in Summer 2015. \textit{Right:} One-dimensional profile of the
  average image shown on the left. The horizontal axis (relative
  counts) is linear. The
  dashed curves are the one-dimensional profile of
  a PSF star taken on the same night (scaled on the horizontal axis to fit on the plot) for
  comparison to the widths of Pluto and Charon. Pluto is
  about 2 PSFs across and so is partially spatially resolved. Charon is
  only slightly wider than the PSF width and is effectively unresolved.}
\end{center}
\end{figure}

\begin{table}[h!]
\begin{center}
\textbf{Table 1}\\
Observational Circumstances\\
\begin{tabular}{ccccccc}
\hline
UT date & Weather & Sub-Earth & Pluto-Charon & Phase &
                                                       \multicolumn{2}{c}{Total exp. time (min)}\\
mean-time & conditions &
                         Lon. ($^{\circ}$E)\renewcommand{\thefootnote}{\fnsymbol{footnote}}\footnote{The
                         rotation periods of Charon and Pluto and
                         Charon's orbital period are equal
                         ($\sim$6.4 days). We use the preferred
                         coordinate system where
                         the sub-Pluto point on Charon is at
                         0$^{\circ}$ longitude and the anti-Pluto point is
                         at 180$^{\circ}$ longitude (Zangari, 2015). The sub-observer longitude
                         decreases with time and differs by
                         180$^{\circ}$ between Pluto and Charon.} &
                                                                    sep. ($''$) & angle ($^{\circ}$) & $H$ band & $K$ band\\
\hline
2015-07-14 10:28 & Clear & 322 & 0.74 & 0.24 & 24 & 30\\
2015-07-18 08:12 & Some cirrus & 102 & 0.84 & 0.36 & 48 & 30\\
2015-07-24 08:11 & Some cirrus & 124 & 0.80 & 0.53 & 48 & 45\\
2015-07-28 10:28 & Clear & 253 & 0.83 & 0.64 & 24 & 65\\
2015-08-08 08:06 & Clear & 359 & 0.65 & 0.94 & 48 & 45\\
2015-08-30 07:55 & Clear & 200 & 0.67 & 1.41 & 48 & 45\\
\hline
\end{tabular}
\end{center}
\end{table}

\section{Reduction}
\onehalfspace{A more detailed description of the reduction
strategy is presented in Appendix A; reduced Charon spectra corrected for
solar and telluric absorption are available as supplementary material
online. The raw data were first processed through the OSIRIS Data
  Reduction Pipeline (DRP; Krabbe et al., 2004) to produce 3D data cubes (1 spectral and 2
  spatial dimensions). We created 2 sets
of data cubes: dark- and sky-subtracted. The master dark of the
appropriate exposure time was subtracted from all 4 images in a set (3
science, 1 sky) to produce the dark-subtracted data cubes. The
sky-subtracted data cubes were constructed by subtracting the sky image of
each set from the other images in the set; dark subtraction was not
explicitly performed because sky subtraction implicitly
includes dark subtraction. The rest of the reduction was handled
using in-house IDL routines. The trace and aperture radii were
determined in each data cube from the mean (spectrally-averaged)
image; different aperture radii were used for
Pluto, Charon, and the solar analog to minimize the addition of
noise. One-dimensional spectra were extracted from the sky-subtracted
data cubes by summing the values in each pixel of the circular apertures
surrounding Pluto, Charon, and the solar analog star. We constructed a wavelength solution from the
measured positions of OH emission lines in the dark-subtracted data cubes and published
vacuum wavelength values (Rousselot et al., 2000); our wavelength
solution was nearly identical to the one output by the
DRP, so we used our wavelength solution. Telluric
absorption was corrected for by dividing the science
spectra by ATRAN models of atmospheric transmission (Lord, 1992). The
telluric-corrected Charon spectra were
divided by the telluric-corrected solar analog spectrum in the
appropriate band from that night to remove solar
absorption features, yielding units of arbitrary albedo. This was
converted to geometric albedo by scaling to the
Charon spectrum in Fig. 3 of Buie and Grundy (2000). An adjustment
to the geometric albedo was applied to correct for differences in radii: Buie and
Grundy (2000) used 593 km, while Stern et al. (2015) report a value of
606 km from New Horizons.}

\begin{figure}[h!]
\begin{center}
\includegraphics[scale=0.75,trim=1.75cm 0.5cm 0cm 1.75cm,clip=true]{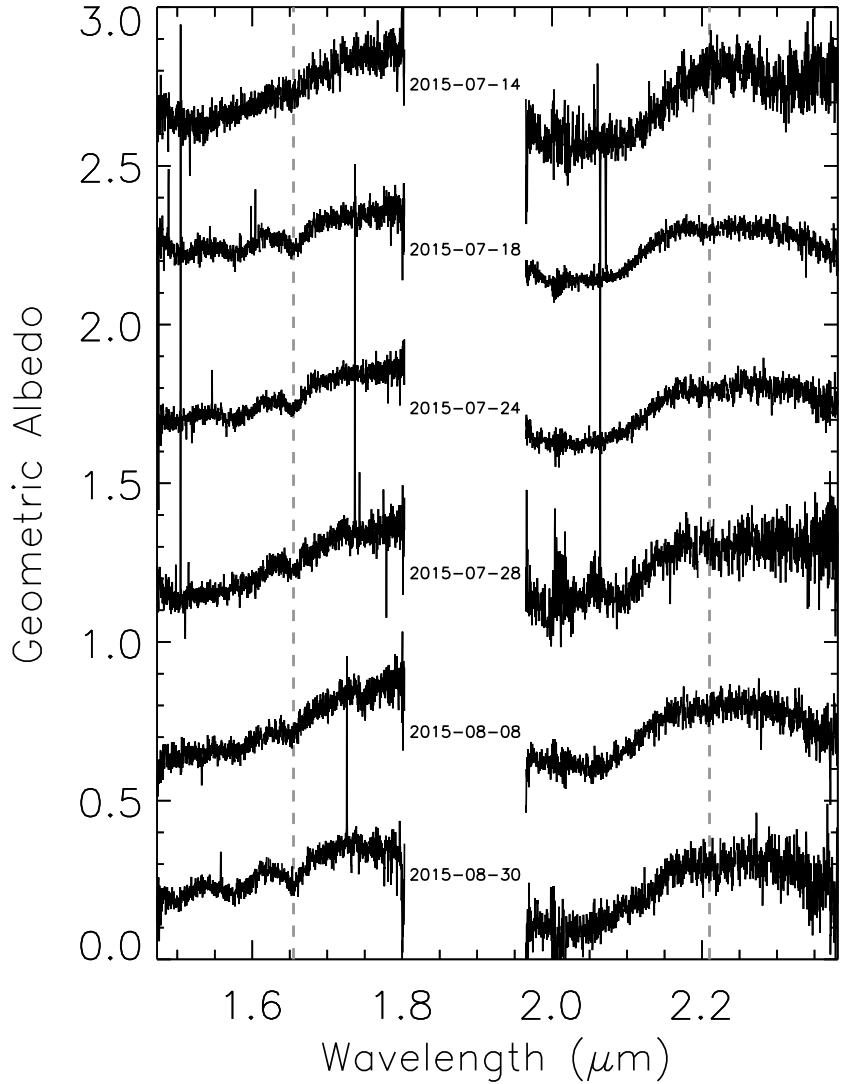}
\caption{Charon night averaged spectra computed as robust averages (3-$\sigma$) of
  the reduced spectra from each given night (dates are marked between
  each $H$ and $K$ pair). The spectra are offset in
  intervals of 0.5 in geometric albedo for clarity. Vertical dashed lines
  mark the approximate centers of the 1.65 $\mu$m crystalline water ice band
  and the 2.21 $\mu$m ammonia hydrate band. The highest signal-to-noise
spectra were obtained on 2015-07-18 and 2015-07-24, and these two
nights dominate the grand average (calculated using a weighted average).}
\end{center}
\end{figure}

\begin{figure}[h!]
\begin{center}
\includegraphics[scale=0.59,trim=0cm 2cm 0cm 3cm,clip=true]{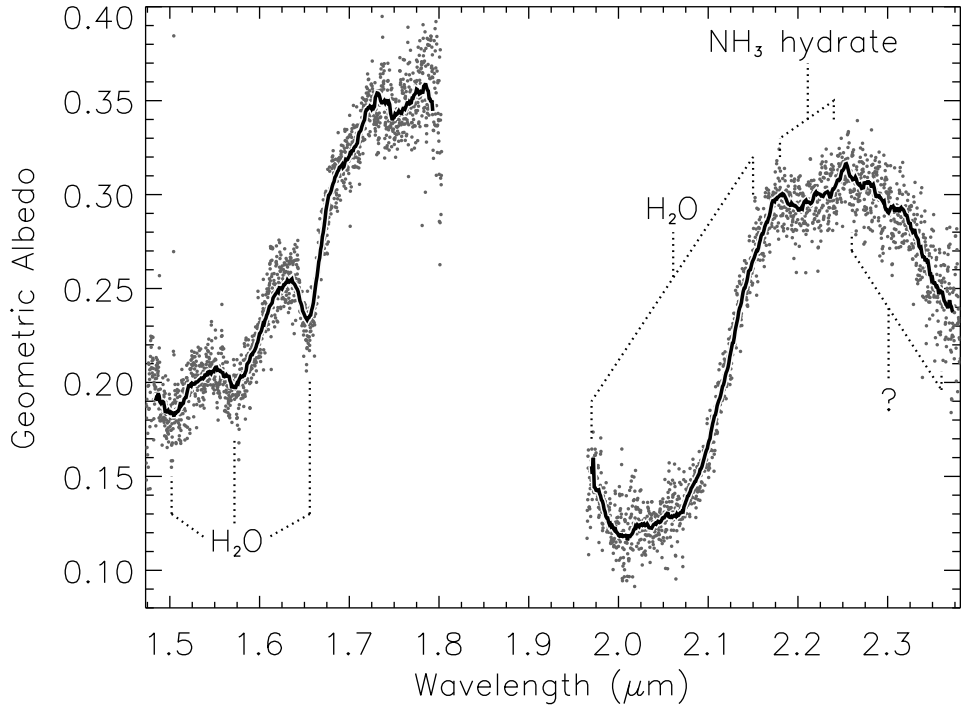}
\caption{Charon grand average spectrum computed as a weighted average
  of the night averaged spectra. This spectrum was
scaled to the Charon spectrum from  3 of Buie and Grundy (2000),
with an additional correction for differences in Charon radii: Buie
and Grundy (2000) used 593 km and we used
606 km from Stern et al. (2015). Exceptionally
strong telluric absorption is present in the blank central portion;
for this reason, no OSIRIS filter covers this wavelength region. The
strong crystalline H$_2$O feature at 1.65 $\mu$m and the weaker ammonia
hydrate feature at 2.21 $\mu$m are the focus of this work. Other
absorption features at 1.5 and 2.0 $\mu$m are due to a combination of amorphous and
crystalline water ice; another crystalline band is present at 1.56
$\mu$m (Mastrapa et al., 2008). The source of absorption at
wavelengths $>$2.25 $\mu$m is unknown (Buie and Grundy, 2000).}
\end{center}
\end{figure}

\section{Analysis}
\onehalfspace{Charon night averaged spectra were calculated as robust (3-$\sigma$) averages of
the individual $H$ and individual $K$ spectra from a given
night. Uncertainties on the albedo in each wavelength bin were
computed as the standard deviation of the albedo values in each bin. The night averaged spectra are presented in
Fig. 3. The Charon grand average spectrum (Fig. 4) was computed as a weighted average
of the night averaged spectra using Gaussian weighting (1/$\sigma^2$).\\
\indent We assumed the 1.65 $\mu$m and 2.21 $\mu$m absorption bands and surrounding continuum
regions had the functional form (Gaussian + 1) $\times$ polynomial. From this we extracted
fit values for the amplitude, band center, and 1/$e$ half width of the
absorption bands. We modeled the
continuum regions (the adjacent regions on either side
of the absorption features) differently for the 1.65 $\mu$m and 2.21 $\mu$m
bands. We performed a linear fit to the relatively linear regions on
either side of the 1.65 $\mu$m band; to avoid influencing the fit, points in the
band itself were not considered. The continuum wavelength regions were 1.610-1.624 $\mu$m and
1.684-1.714 $\mu$m. For the 2.21
$\mu$m band, we fit a third order polynomial to the wavelength
regions between 2.115-2.190 $\mu$m and 2.230-2.320 $\mu$m. The entire spectrum was
then divided by the best-fit polynomial, setting the continuum regions to an
average value of +1. This offset was removed so that the continuum
was at an average value of zero, thus removing a
parameter from the Gaussian fit to the band. Using the IDL
routine
\textit{mpfitpeak}\renewcommand{\thefootnote}{\fnsymbol{footnote}}\footnote{The
  IDL routine \textit{mpfitpeak} fits a Gaussian, Lorentzian, or
  Moffat function to data. Documentation
  for \textit{mpfitpeak} can be found at \url{http://hesperia.gsfc.nasa.gov/ssw/gen/idl/fitting/mpfit/mpfitpeak.pro}.}, a Gaussian of the form
$ae^{-(\lambda-\mu)^2/2w^2}$ was fit from 1.610-1.714 $\mu$m
for the 1.65 $\mu$m band and from 2.190-2.230 $\mu$m for the 2.21 $\mu$m
band. The band depth (amplitude), $a$, band center, $\mu$, and 1/$e$
half width, $w$, were the fit parameters. We fit for the 1/$e$ half
width, however, the full width at half maximum (FWHM) can be
calculated from this quantity by means of
FWHM=2$w\sqrt{2\mathrm{ln}2}$. The band area, a quantity related to
equivalent width, provides a measure of the amount of an absorbing
species on the surface and is proportional to the product of
the band depth and the 1/$e$ half width ($aw\sqrt{2\pi}$). The band
center was used to calculate the ice temperature from the 1.65 $\mu$m
crystalline water ice band, as described later in this section. Uncertainties on the fit parameters
were determined using standard methods for least squares
fitting (e.g., Section 15.4 of Press et al., 2007).

\begin{figure}[h!]
\begin{center}
\includegraphics[scale=0.65,trim=0.25cm 0.75cm 0cm 1.25cm,clip=true]{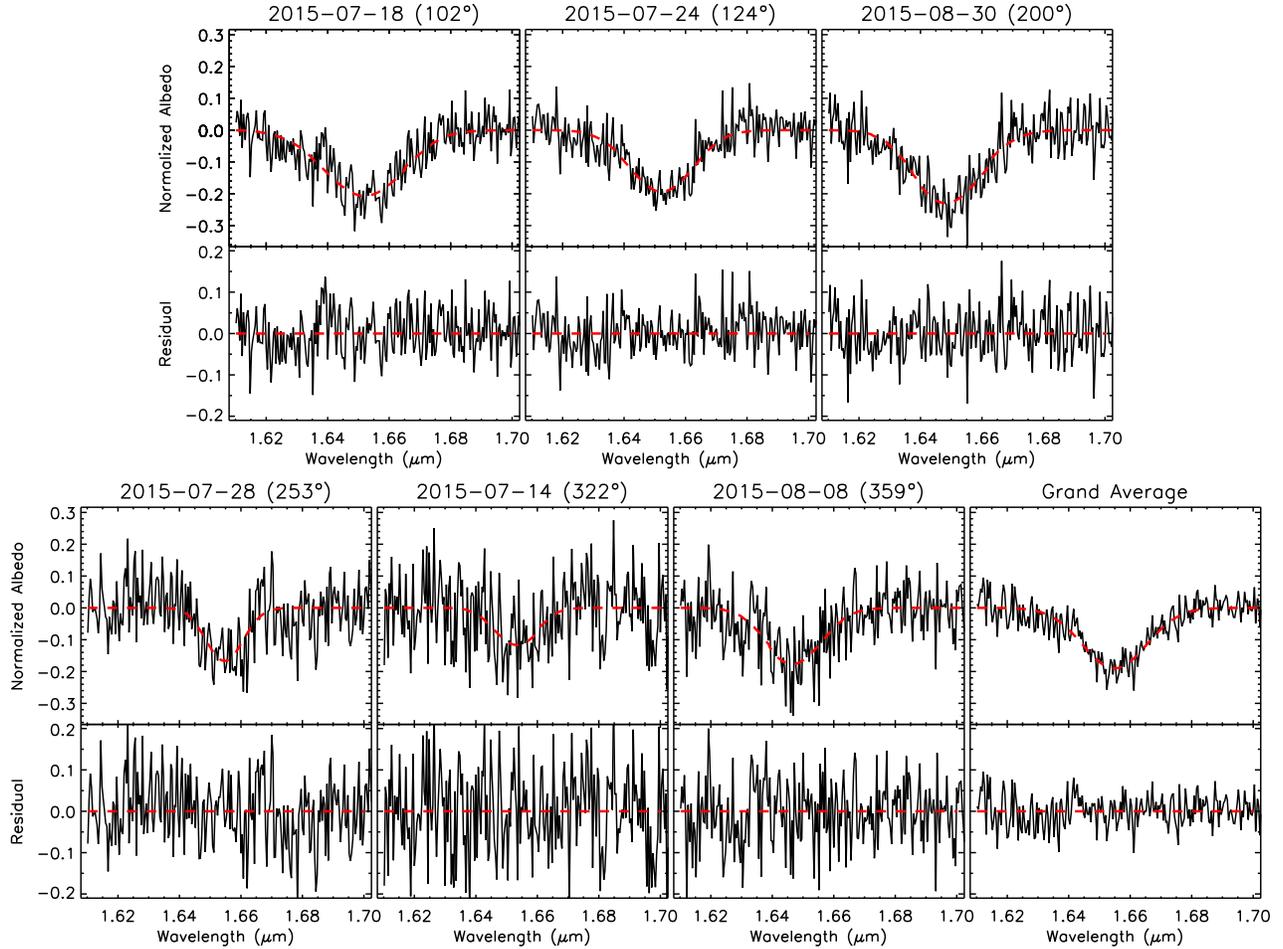}
\caption{\textit{Top panels in each row:} Gaussian fits (dashed line) to the data
  for the 1.65 $\mu$m crystalline H$_2$O band. The continuum on either
  side of the band is centered at zero (see text for
  explanation). \textit{Bottom panels in each row:} Residuals (data
  minus model) scattered about zero (dashed line).}
\end{center}
\end{figure}

\begin{figure}[h!]
\begin{center}
\includegraphics[scale=0.65,trim=0.25cm 0.75cm 0cm 1.25cm,clip=true]{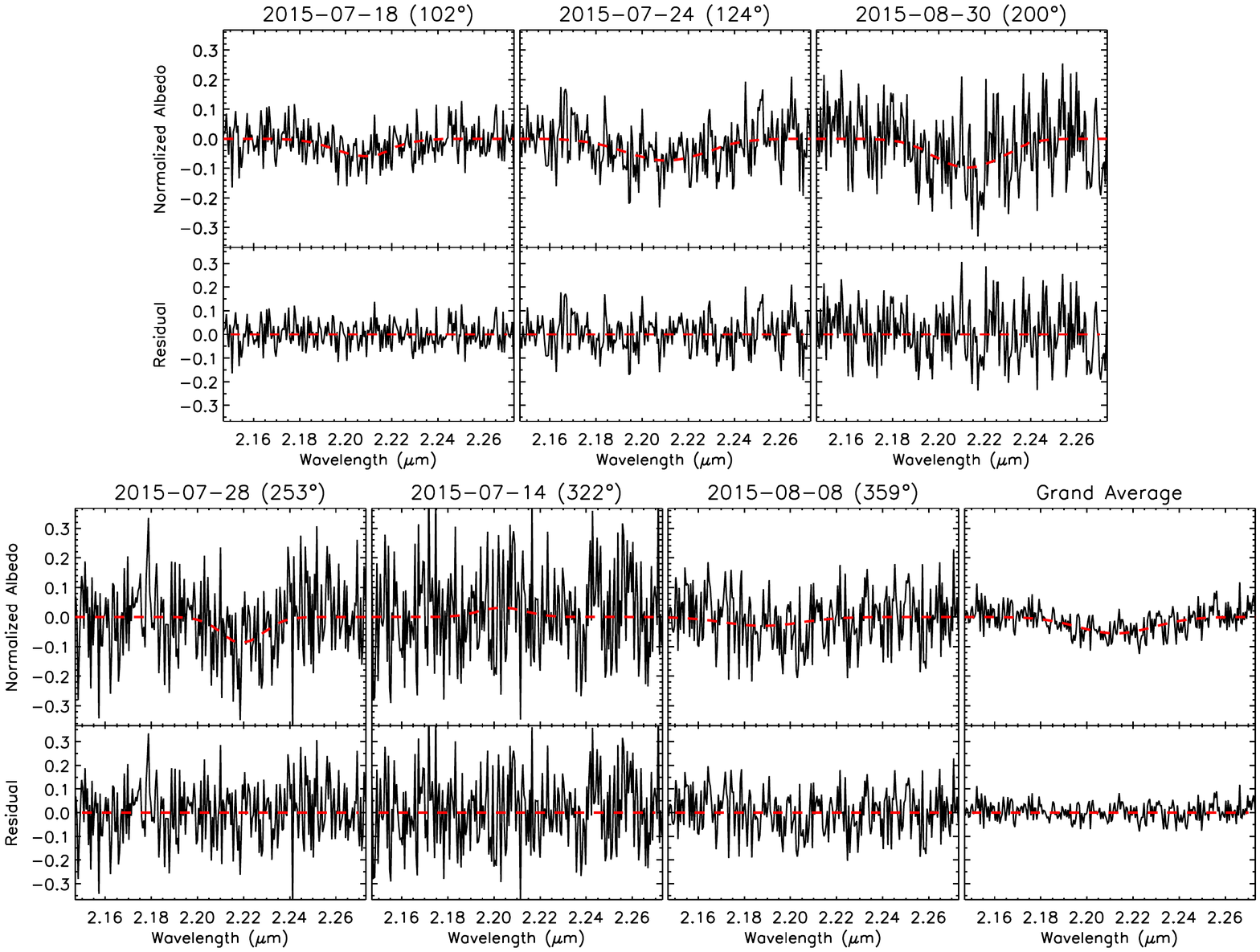}
\caption{\textit{Top panels in each row:} Gaussian fits (dashed line) to the data
  for the 2.21 $\mu$m ammonia hydrate band. The continuum on either
  side of the band is centered at zero (see text for
  explanation). \textit{Bottom panels in each row:} Residuals (data
  minus model) scattered about zero (dashed line).}
\end{center}
\end{figure}

\indent As a side note, we initially
performed a full 5-parameter sloped Gaussian fit to both absorption bands, as was
done in DeMeo et al. (2015) for the 2.21 $\mu$m band. The resulting
fits poorly matched the data and were accompanied by
large uncertainties on the parameters. Fitting so many parameters at
once was likely the cause of the poor fits. The two-step process described above was significantly more
effective at modeling the absorption features. Additionally, we found
that a third order polynomial was a better fit to the continuum on
either side of the 2.21 $\mu$m feature, and used that as our model
instead of a line.\\
\indent Fig. 5 and 6 show the Gaussian fits to the data and residuals (data minus model) for the
1.65 $\mu$m and 2.21 $\mu$m bands, respectively. In Tables 2 and 3, the
values reported in parentheses are uncertainties on the last two
significant figures of the parameter value. Additionally, Table 2 provides values and uncertainties
for the wavenumber of the band center, $\bar{\nu}=1/\lambda$. Table 3
also presents the FWHM and the band area. Values from the
grand average spectrum (Fig. 4) are also found in Tables 2 and 3 (see
``Grand avg.'' rows); values and uncertainties from the grand average are
presented in Fig. 7 and 9 as solid and dashed lines, respectively. The
values from the grand average spectrum are in agreement with the
averages of all the nightly values, within the uncertainties. We
ignored phase angle effects in this work and the temporal evolution of surface ice distributions on
Charon due to the short duration over which the observations were obtained ($\sim$1.5 months).\\
\indent A look-up table, created using the IDL routine
\textit{alpha\_h2o}\renewcommand{\thefootnote}{\fnsymbol{footnote}}\footnote{\url{http://www2.lowell.edu/~grundy/abstracts/ice/alpha_H2O.pro}}
and data from Grundy and Schmitt
(1998), was used to determine the temperatures that correspond to the
calculated 1.65 $\mu$m band centers. The \textit{alpha\_h2o} routine
computes the absorption coefficients, $\alpha$, of crystalline water ice as a
function of wavelength between 20 and 270 K over a
wavelength range of the user's choice. We used
\textit{alpha\_h2o} to compute the absorption coefficients for
temperatures between 20 and 100 K in steps of
1 K from 1.6102-1.7142 $\mu$m. The location of the 1.65 $\mu$m band center at each
temperature value was obtained (to a
precision of 0.0001 $\mu$m) by determining the wavelength
corresponding to the maximum absorption coefficient in
the modeled wavelength range (a larger absorption coefficient results in a
deeper absorption band in the near-infrared spectrum). These
temperature-wavelength pairs were placed in our look-up table and used
to determine the surface ice temperatures and uncertainties presented
in Table 2.}

\begin{table}[h!]
\begin{center}
\textbf{Table 2}\\
Gaussian Fit Parameters: 1.65 $\mu$m H$_2$O Band \\
\begin{tabular}{ccccccc}
\hline
UT date & $a$ & $\mu$ ($\mu$m) & $w$ ($\mu$m) & $\bar{\nu}$ (cm$^{-1}$) & $T$ (K)\\
\hline
2015-07-14 & -0.117(67) & 1.6562(48) & 0.0073(52) & 6038$\pm$18 & 28$^{+60}_{-28}$\\
2015-07-18 & -0.207(23) & 1.6555(18) & 0.0132(21) & 6040.3$\pm$6.4 & 37$\pm$24\\
2015-07-24 & -0.194(25) & 1.6543(16) & 0.0105(18) & 6044.9$\pm$5.7 & 53$\pm$21\\
2015-07-28 & -0.167(53) & 1.6555(22) & 0.0061(23) & 6040.5$\pm$8.0 & 37$\pm$29\\
2015-08-08 & -0.177(45) & 1.6526(29) & 0.0096(32) & 6051$\pm$10 & 74$\pm$37\\
2015-08-30 & -0.230(25) & 1.6541(16) & 0.0120(18) & 6045.4$\pm$5.7 & 56$\pm$21\\
Grand avg. & -0.190(17) & 1.6549(11) & 0.0105(12) & 6042.6$\pm$4.0 & 45$\pm$14\\
\hline
\end{tabular}
\end{center}
\end{table}

\begin{table}[h!]
\begin{center}
\textbf{Table 3}\\
Gaussian Fit Parameters: 2.21 $\mu$m Ammonia Hydrate Band\\
\begin{tabular}{ccccccc}
\hline
UT date & $a$ & $\mu$ ($\mu$m) & $w$ ($\mu$m) & FWHM ($\mu$m) & Band Area ($\mu$m)\\
\hline
2015-07-14\renewcommand{\thefootnote}{\fnsymbol{footnote}}\footnote{The
  low SNR of the night averaged spectrum for 2015-07-14 UT is likely
  responsible for the positive amplitude of the Gaussian fit to the
  2.21 $\mu$m band on this night (Fig. 6).} & 0.031(91) & 2.207(34) & 0.011$^{+0.039}_{-0.011}$ & 0.025$^{+0.091}_{-0.025}$ & 0.0008$^{+0.0039}_{-0.0008}$\\
2015-07-18 & -0.059(26) & 2.2099(64) & 0.0133(77) & 0.031(18) & 0.0020(14)\\
2015-07-24 & -0.074(34) & 2.2118(83) & 0.018(11) & 0.042(25) & 0.0033(25)\\
2015-07-28 & -0.086(64) & 2.2261(83) & 0.0101(96) & 0.024(23) & 0.0022$^{+0.0026}_{-0.0022}$\\
2015-08-08 & -0.030(63) & 2.190(28) & 0.018$^{+0.043}_{-0.018}$ & 0.04$^{+0.10}_{-0.04}$ & 0.0013$^{+0.0043}_{-0.0013}$\\
2015-08-30 & -0.097(54) & 2.2167(92) & 0.015(11) & 0.036(27) & 0.0038(35)\\
Grand avg. & -0.055(18) & 2.2125(56) & 0.0164(71) & 0.039(17) & 0.0023(12)\\
\hline
\end{tabular}
\end{center}
\end{table}

\section{Results \& discussion}
\subsection{Water ice temperature}
\onehalfspace{The results of the ice temperature calculation from the
  shift of the 1.65 $\mu$m crystalline H$_2$O band are presented in Fig. 7. The
  mean surface ice temperature on the observable portion of Charon and
  its uncertainty, 45$\pm$14 K, are represented by horizontal lines. No longitudinal
  temperature variations were detected on Charon and all temperature measurements are
  consistent with the grand average value, as expected. The large uncertainties on temperature in this work
  are due to a combination of modeling uncertainties in Grundy and
  Schmitt (1998) and the SNR of the 1D spectra in this work. Summing
  all the pixels of the Charon aperture within each wavelength bin
  introduces read noise from each pixel (the Charon aperture was typically
  16 pixels in area). Our observations are read noise limited so
  this is a significant contributor to the noise in the final 1D spectra.\\
\indent Our temperature distribution agrees with the results of Cook et al. (2007), who
  report values of 42.5$\pm$10 K on the sub-Pluto hemisphere (270$^{\circ}$-90$^{\circ}$) and
  52.7$\pm$10 K on the anti-Pluto hemisphere
  (90$^{\circ}$-270$^{\circ}$). Averaging the values from Cook et
  al. (2007) yields a mean temperature of
  47.6$\pm$7.1 K, which is in good agreement with our value of
  45$\pm$14 K. Our mean ice temperature also agrees
  with the mean surface temperature of 55.4$\pm$2.6 K from Spitzer reported by
  Lellouch et al. (2011). Comparing our mean ice temperature to thermal measurements and
thermophysical modeling, we note that long-wavelength thermal emission
is proportional to temperature (Rayleigh-Jeans approximation) so the
brightness temperature of
43.7$\pm$0.2 K from ALMA (Bryan Butler, personal communication) is
directly comparable to, and in good agreement with, our measurement of
45$\pm$14 K. Pluto and Charon were
  unresolved in the Spitzer observations but fully separated in the
  ALMA observations.\\
\indent We investigated if this
  temperature distribution was consistent with the observed
  visible geometric albedo ($p_V$) variations across Charon (Buie et al.,
  2010). For this calculation, we assumed that the ice temperature was proportional to
  (1-$A$)$^{1/4}$, where the bolometric Bond albedo, $A$, is the product of $p_V$
  and the phase integral, $q$. Taking the global geometric albedo to
  be 0.41 (Buratti et al., submitted) and using a reasonable assumption for the
  form of the phase integral ($q$=0.336$p_V$+0.479; Brucker et al.,
  2009), we calculate a Bond albedo of 0.25, matching the value
  reported by Buratti et al. The 8\% variation
  in geometric albedo produces a temperature variation of $<$1 K across the
  surface of Charon. From this work, the level of precision on the
  temperature measurements is not sufficient to rule out these small temperature variations.

\begin{figure}[h!]
\begin{center}
\includegraphics[scale=0.75,trim=2cm 10cm 0cm 1.75cm,clip=true]{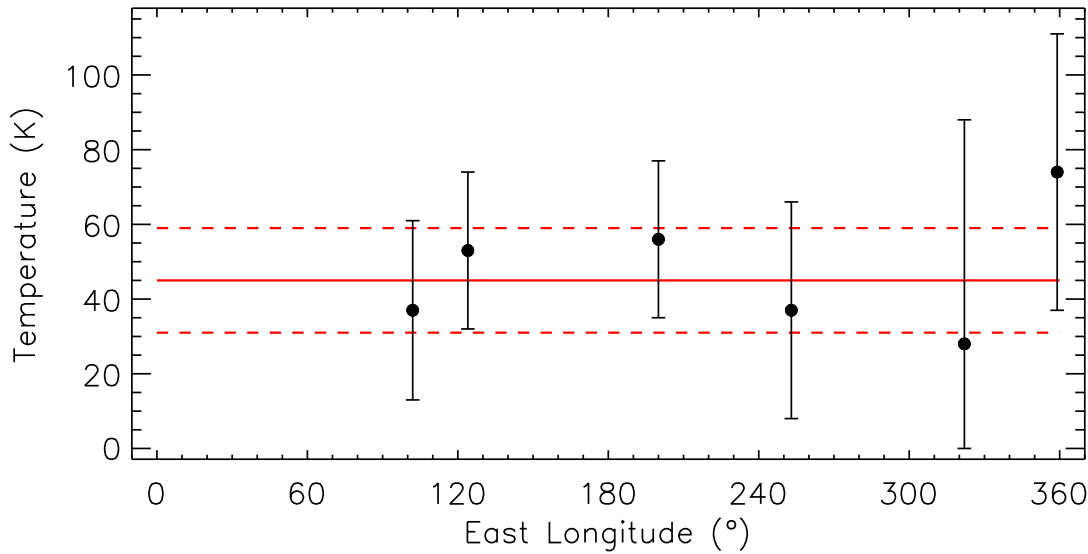}
\caption{Variation of ice temperature with sub-observer longitude. The solid line represents the
  mean surface ice temperature on the observable portion of Charon calculated from the grand
  average spectrum. The dashed lines and error bars are
  1-$\sigma$ uncertainties. Temperature values are truncated at 0 K,
  resulting in asymmetric error bars for the point at 322$^{\circ}$
  longitude. All night averaged temperature values are
  consistent with the globally-averaged temperature of 45$\pm$14 K.}
\end{center}
\end{figure}

\subsection{Water ice phase}
The presence of crystalline water ice on Charon is not in
doubt since the 1.65 $\mu$m band, unique to the crystalline phase, is
clearly detected in spectra. What is less obvious is the reason why crystalline water ice is
present at all, not to mention that it is present in such large quantities ($\sim$90\%;
Cook et al., 2007; Merlin et al., 2010). Conversion of water ice from
the crystalline phase to the amorphous phase should take $\sim$1.5
Myr (Cooper et al., 2003), assuming a radiation environment of 1
eV-10 GeV protons and no recrystallization
processes. Other authors have suggested that the large quantity of
crystalline water ice is due to surface replenishment processes such
as cryovolcanism or solid state convection (e.g.: Cook et al., 2007; DeMeo
et al., 2015). Due to Charon's small diameter, low density ($\sim$1.7 g
cm$^{-3}$), highly circular orbit ($e$=0.00005),
and ancient surface ($\geq$4 Gyr), the most likely sources that would
drive these processes, internal and tidal heating, are probably
negligible in the present day (Stern et al., 2015). Indeed, no signs of
present-day cryovolcanic or solid state convection were noted on
Charon's encounter hemisphere (Moore et al., 2016).\\
\indent Laboratory studies of water ice in its various phases provide
a less complicated explanation. Conversion of water ice from
the crystalline to the amorphous phase is a reversible process that
depends on temperature (Leto and Baratta, 2003; Mastrapa and Brown,
2006; Zheng et al., 2009). These studies used different particles
(Lyman-$\alpha$ photons, ions, and electrons, respectively) to
bombard samples of crystalline water ice, but the results are
comparable since the secondary processes (interactions with electrons
and ions produced during previous reactions) are independent of the
primary particle and are more important than the identity of the primary
particle in altering the state of the water ice (Zheng et al.,
2009). The conclusion of these
experiments is that thermal recrystallization provides a non-negligible
balance to irradiation amorphization at temperatures greater than 30
K. After an appropriate amount of time has elapsed (dependent on the
temperature) an equilibrium is reached
between these two processes and the crystalline-to-amorphous ratio
remains constant. Fig. 8 shows that the time needed to reach this
equilibrium is less than 1 Gyr for temperatures between 30 and 50
K. Also, the equilibrium ratio of crystalline-to-amorphous water ice
is higher at higher temperatures. Zheng et al. (2009) did not
quantify the crystalline-to-amorphous ratio explicitly, but instead
measured a related quantity, $\delta$, the ratio of the 1.65 $\mu$m band depth after irradiation to
the band depth prior to irradiation. We used Eq. 1 from Zheng et al. (2009),
$\delta$($t$)=1-$B$(1-$e^{-t/k_1}$), where the constants $B$ and $k_1$ are
amplitude and $e$-folding time, respectively, to construct the curves
in Fig. 8. We took the values for $B$ and $k_1$ at
30 K ($B$=0.79$\pm$0.06, $k_1$=0.109$\pm$0.018 Gyr) and 50 K
($B$=0.39$\pm$0.06, $k_1$=0.087$\pm$0.018 Gyr) from Table 1 of Zheng et
al. (2009). The scale factor between irradiation
time in the laboratory and actual time on the surface of a KBO is 44
hours per 1.6 Gyr (Cooper et al., 2003; Zheng et al., 2009).\\
\indent Since thermal recrystallization rates increase with temperature (Zheng
et al., 2009), the equilibrium state between crystalline and amorphous
water ice will be weighted more toward the maximum temperature
reached. We consider Charon as a fast and slow rotator to
calculate the range of maximum temperatures. A fast rotator is an
object with a high thermal inertia, meaning that the surface will
maintain an elevated temperature for a considerable period following
the time of maximum solar insolation (local noon). A slow rotator has a low thermal inertia, resulting in the
entirety of the surface in instantaneous equilibrium with incoming solar
radiation. For a fast rotator,
(e.g., Sicardy et al., 2011), this is near 50-53 K assuming $A$$\sim$0.25 as
before, and a beaming parameter of 0.9-0.7. (The beaming parameter
takes into account the increased flux from an object at small phase
angles, a phenomenon known as the opposition effect.) For Charon as a
slow rotator, the maximum temperature
would be much higher, 67-71 K, beyond the range of temperatures probed
by Zheng et al. (2009). Lellouch et al. (2011) find that Charon is not
in instantaneous equilibrium with incoming solar radiation and is therefore not
a slow rotator.\\
\indent Moore et al. (2016) indicate that the youngest surface age on
Charon is $\sim$4 Gyr, suggesting that placement of crystalline water
ice occurred at this time. It is likely that Charon underwent differentiation
following the giant impact that created the system (Canup, 2011) and
that water, in some form, was brought to the surface
afterwards. Extensional features observed on Charon may be due to a
sub-surface ocean that froze, resulting in the eruption of ice onto the surface
(Moore et al., 2016). This cryovolcanic period $\sim$4 Gyr
ago was likely global in extent since
water ice absorption is observed at every longitude on Charon (Grundy et al., 2016).

\begin{figure}[h!]
\begin{center}
\includegraphics[scale=0.5,trim=0cm 2cm 0cm 3.5cm,clip=true]{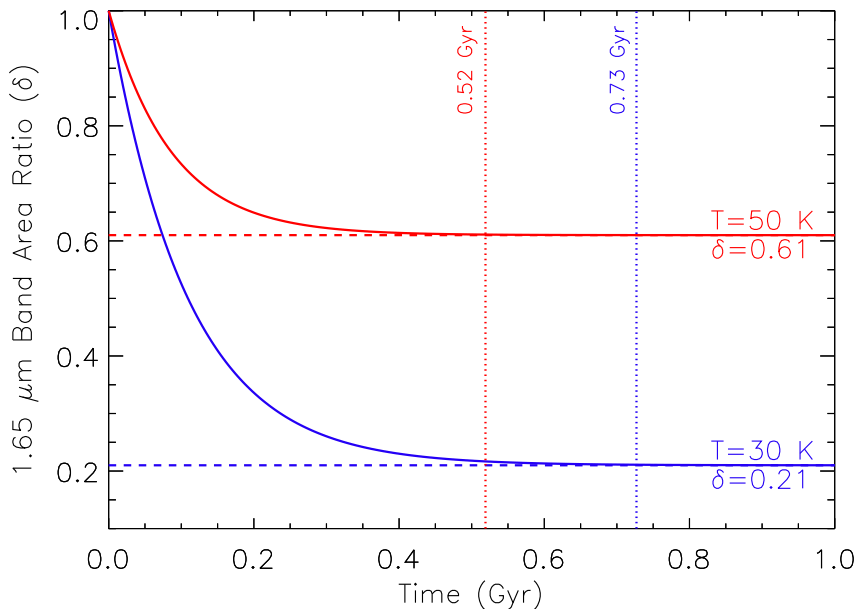}
\caption{Ratio of the band area after irradiation to the original band
  area of the 1.65 $\mu$m crystalline water ice absorption feature
  plotted against time for surface temperatures of 30 K (lower curve) and 50
  K (upper curve). See the text for the equation for $\delta$; constants were
  taken from Table 1 of Zheng et al. (2009). The
  values of $\delta$ presented on the right are the asymptotic values
  that the curves approach. The vertical dotted lines mark the
  point in time when the slope has effectively flattened out,
  corresponding to approximately 6.5$k_1$, where $k_1$ is the
  $e$-folding time of $\delta$ (time
  required for $\delta$ to decrease to 1/$e$ of its original
  value when the sample is exposed to radiation).}
\end{center}
\end{figure}

\indent Crystalline water ice was detected on the other large KBOs Haumea (Trujillo et
al., 2007; Dumas et al., 2011), Orcus (de Bergh et al., 2005), and
Quaoar (Fornasier et al., 2004; Jewitt and Luu, 2004). A
replenishment mechanism has been invoked to explain the presence of
crystalline water ice on each of these bodies. The surface
temperatures of these objects are: $\leq$40 K for Haumea (Merlin et
al., 2007), $\leq$44 K for Orcus (Barucci et al., 2008), and 44 K for Quaoar
(Fraser et al., 2013). It is conceivable
that thermal recrystallization is an important process on these KBOs
as well. Improved temperature estimates for these bodies would provide
vital tests of this theory. Firm detections of
crystalline water ice on smaller objects where surface replenishment
is unlikely, particularly the four minor moons
of Pluto (Styx, Nix, Kerberos, and Hydra) supports this theory (Cook et al., 2016).

\subsection{Ammonia hydrate ice}
\indent The results of the band-fitting analysis for the 2.21 $\mu$m
ammonia hydrate absorption band are presented in Fig. 9. Four separate plots are
presented: band center, band depth, band FWHM, and band area
($aw\sqrt{2\pi}$) as functions of sub-observer longitude. Grand average values are again
presented as horizontal lines in each plot. All quantities are
consistent with their respective grand average values, as
expected. Note that the plotted points indicate the sub-observer
longitude at the time of the observations but the entire visible
hemisphere of Charon (sub-observer longitude $\pm$ 90$^{\circ}$)
contributes flux to the spectra. We report the detection of ammonia
hydrate on the visible hemispheres centered on the sub-observer
longitudes 102$^{\circ}$, 124$^{\circ}$, and 200$^{\circ}$,
corresponding to a total detection range of 12$^{\circ}$ to
290$^{\circ}$ longitude. Non-detections of ammonia hydrate are
reported at the visible hemispheres centered on the sub-observer
longitudes 253$^{\circ}$, 322$^{\circ}$, and 359$^{\circ}$ (the error
bars on the band areas at these sub-observer
longitudes, shown in the bottom plot of Fig. 9, extend to 0 $\mu$m,
suggesting the possibility of no absorption due to ammonia hydrate).\\
\indent\indent Fig. 3 from DeMeo et al. (2015) presents the same
ammonia hydrate band properties plotted against longitude (except that our band area
includes an extra factor of $\sqrt{2\pi}$), and they report statistically
significant variations in band center and band depth. We report no
statistically significant variations in any of the ammonia hydrate
band properties; the higher SNR
of their data may be responsible for the discrepancy. Grundy et al. (2016)
reported a low level of ammonia at every longitude of the New Horizons
encounter hemisphere ($\sim$270-60$^{\circ}$), with higher concentrations in bright-rayed
craters (see Fig. 8c of that paper). Dalle Ore et al. (2016) report uniform distribution
and composition of ammonia species across the equatorial region of the
encounter hemisphere from New Horizons data. In terms of longitudinal
coverage, our results complement the ammonia
distribution from New Horizons (Dalle Ore
et al., 2016; Grundy et al., 2016), and together they suggest that
ammonia is uniformly distributed across the surface of Charon.

\begin{figure}[h!]
\begin{center}
\includegraphics[scale=0.65,trim=0cm 1.0cm 0cm 1.8cm,clip=true]{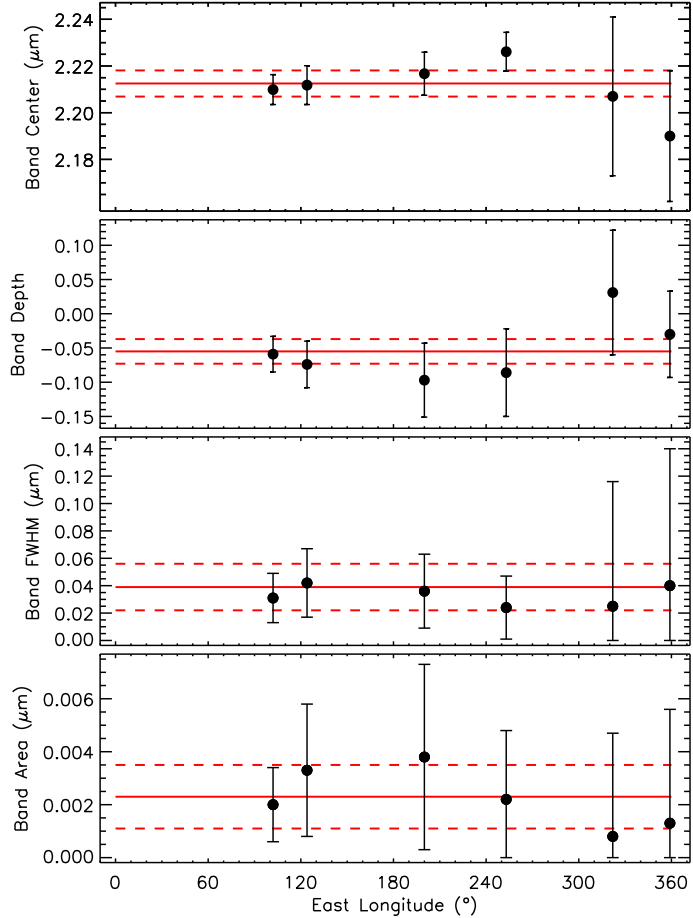}
\caption{In all panels, the value obtained from fitting the 2.21 $\mu$m ammonia hydrate
  band from the grand average spectrum (Fig. 6) is
  shown as a solid horizontal line. The dashed lines and error
  bars represent 1-$\sigma$ uncertainties. The values for FWHM and
  band area are truncated at 0 $\mu$m, resulting in asymmetric error bars
  for some points (Table 3). All quantities are consistent with their
  respective grand average values. \textit{Top panel:}
  Variation of the 2.21 $\mu$m band center with sub-observer
  longitude. \textit{Top middle panel:} Variation of the 2.21 $\mu$m band depth
  (measured in normalized albedo) with sub-observer
  longitude. This band depth is measured as the difference between
  zero and the minimum of the Gaussian, resulting in a negative
  number. Positive values simply mean
  that the spectrum was noisy enough to increase the uncertainty on
  the fit parameters or prevent detection of the absorption
  feature altogether. \textit{Bottom middle panel:} Variation of the 2.21
  $\mu$m band full width at half maximum (FWHM) with sub-observer
  longitude. \textit{Bottom panel:} Variation of the 2.21 $\mu$m band
  area with sub-observer longitude. The band area was calculated as
 $aw\sqrt{2\pi}$, yielding units of $\mu$m.}
\end{center}
\end{figure}

\indent The ratio of pure ammonia to its hydrated states is unknown and
we provide no additional constraints because of the difficulty in separating the
effects of these different species on the shape of the 2.21 $\mu$m band. Ammonia and its
various hydrated states absorb at this wavelength, with the hydrates
broadening the band (Strazzulla and Palumbo, 1998; Moore et al.,
2007); the grand average and most night averaged spectra show a
broad absorption feature, requiring the presence of hydrated
ammonia (Fig. 3, 4, and 6). The 2.21 $\mu$m band depth and band area distributions show
no variability, within the error bars, indicating a uniform
concentration and/or grain size. Statistically
significant differences in band center across Charon's surface would
suggest regional differences in the mixture of ammonia and ammonia
hydrates (Strazzulla and Palumbo, 1998; Moore et al., 2007). Since no
variation in band center was detected at the spatial resolution of this
work, the mixture of ammonia species is assumed to be uniform as well.\\
\indent The presence of ammonia on Charon is unexpected at first
given that a majority of ammonia in the upper ice layers should be
destroyed within 20 Myr in a 1 eV-10 GeV proton radiation
environment (Cooper et al., 2003; Cook et al.,
2007). For crystalline water ice, radiation breaks and reforms the
bonds between molecules, thus converting water ice from the crystalline to the amorphous
phase. For ammonia ice, the individual molecules are dissociated by
radiation and cosmic rays, not undergoing a phase conversion. Although
ammonia can also exist in crystalline and amorphous phases,
absorption at 2.21 $\mu$m is not unique to one phase (Moore et al.,
2007). In pure ammonia ice, reforming dissociated ammonia molecules is possible, however, the
presence of water molecules hinders this process, and is
likely the case on Charon's water ice dominated surface (Moore et al.,
2007). Some process is likely renewing ammonia in the surface layers by drawing on a reservoir that is
protected from radiation and cosmic rays.\\
\indent One possibility is the diffusion of ammonia through
a thick layer of water ice, a process that also results in the
hydration of ammonia (Uras and Devlin, 2000; Livingston et al.,
2002). The ammonia
may be primordial, contained in the progenitor
bodies that collided to form Pluto and Charon (Canup,
2011). Ammonia depresses the freezing point of
water and potentially played a role in the cryovolcanic episode
$\sim$4 Gyr ago (Moore et al., 2016). The ammonia in the upper layers
would then have been quickly destroyed, but the ammonia further from
the surface would have survived. Ammonia on Charon's surface in the
present day can be explained by the slow
diffusion rate through the upper layers of water ice and by impact
gardening. This idea is supported by Fig. 8c from
Grundy et al. (2016), which shows a high concentration of ammonia ice
correlated with the bright rays of an impact crater, suggesting that
sub-surface ammonia was excavated by the impact. Additionally, the heat of
the impact may have increased the diffusion rate of ammonia in this localized
region. Not all craters with bright rays have higher concentrations of ammonia though,
likely because of their different times of formation and the
relatively fast dissociation of ammonia molecules. The bright rays would not have
enough time to darken prior to the destruction
of the ammonia.\\
\indent Ammonia was positively identified on Orcus
(Barucci et al., 2008) and the minor moons of Pluto (Cook et al.,
2016). Ammonia was tentatively identified on
Quaoar (Barucci et al., 2015). Observations of Orcus at different sub-observer longitudes
would provide information on the distribution of ammonia across the surface of
the KBO. Barucci et al. (2008) only observed Orcus for 100 minutes, or about
8\% of its rotation period (Rabinowitz et al., 2007), so the spectra
are only from one hemisphere. Ammonia may be present on Haumea, a
KBO with very strong water ice absorption bands, and its largest moon,
Hi'iaka (Barkume et al., 2006). Others argue that ammonia is absent on
Haumea (Trujillo et al., 2007; Pinilla-Alonso et al., 2009). Higher SNR spectral observations
of Haumea are necessary to determine if ammonia is present at
any sub-observer longitude. Haumea also underwent a giant
impact at some point in its history, one large enough to
remove ice, spin-up the body into an ellipsoid, form two satellites,
and create a collisional family in the Kuiper Belt (Brown et al.,
2007). Since they all have similar origins, comparison of the surface compositions of Charon, Haumea, and
its family members may therefore provide insight into the origin of
ammonia on KBOs. In general, further spectral
observations of small to intermediate-sized KBOs, even if ammonia
is not detected, would improve our understanding of the surface
processes at work on these bodies.}

\section{Summary}
\onehalfspace{Six nights of near-infrared spectral observations
  of Pluto and Charon were obtained at different sub-observer longitudes
  between July 14 and August 30, 2015 UT, with the OSIRIS instrument on
  Keck I. These ground-based spectra have a higher spectral
  resolution than the LEISA instrument on New Horizons (Reuter et al.,
  2008; Young et al., 2008), complementing the high spatial resolution data obtained by New
  Horizons. Charon spectra, uncontaminated by reflected
  light from Pluto, were extracted from our
  data. We analyzed the 1.65 $\mu$m crystalline water ice and
  the 2.21 $\mu$m ammonia hydrate ice absorption features with the results
  summarized below:
\begin{itemize}
\item The ice temperature on Charon was calculated from the band
  center shift of the temperature-dependent 1.65 $\mu$m crystalline
  H$_2$O band (Grundy and Schmitt, 1998). The mean surface ice
  temperature on the observable portion of Charon, calculated
  from the grand average spectrum, is 45$\pm$14 K. This is consistent
  with the results of previous work (Cook et al., 2007; Lellouch et al., 2011;
  Bryan Butler, personal communication). The temperature as
  a function of longitude shows negligible variation across the
  surface of Charon and small temperature variations ($<$1 K) due
  purely to albedo variations cannot be ruled out at our precision.
\item At temperatures $\geq$30 K, complete amorphization of
  crystalline water ice does not occur (Leto and Baratta, 2003;
  Mastrapa and Brown, 2006; Zheng et al., 2009). An equilibrium is
  reached between irradiation amorphization and thermal
  recrystallization, and therefore between the crystalline and amorphous water ice
  phases, between 0.52 and 0.73 Gyr for a surface
  temperature between 30 and 50 K (Zheng et al., 2009). The maximum
  surface temperature is likely higher, about 50-53 K for a
  fast-rotating Charon  at its current
  distance from the Sun. The placement of crystalline water
  ice on the surface of Charon likely occurred $\sim$4 Gyr ago (Moore et
  al., 2016), meaning that equilibrium was reached over $\sim$3.5
  Gyr ago. We do not believe that cryovolcanism is necessary to explain
  the presence of crystalline water ice on the surface of Charon.
\item Ammonia hydrate was detected between 12$^{\circ}$ and 290$^{\circ}$
  longitude in this work. In agreement with results from New Horizons
  (Grundy et al., 2016; Dalle Ore et al., 2016), we find that ammonia
  species on Charon are globally distributed. The longitudinal distributions
  of the band center, band depth, FWHM, and band
  area of the 2.21 $\mu$m ammonia hydrate absorption feature were
  found to be uniform. The lack of variability of the band center
  points to a uniform composition of ammonia species across the
  surface.
\item The presence of ammonia ice everywhere on Charon's surface
  requires a means of replenishment since it is dissociated on $\sim$20
  Myr timescales (Cooper et al., 2003; Cook et al., 2007). Ammonia was
  likely a primordial component of the progenitor bodies that collided
  to form Pluto and Charon, and may have played a role in the global
  cryovolcanic episode $\sim$4 Gyr ago (Moore et al., 2016). Since that time, ammonia has been
  slowly diffusing its way through the thick overlying layer of water
  ice and may occasionally be brought to the surface in larger quantities by impacts. This explains the
  presence of hydrated ammonia and its ubiquity across the surface.
\end{itemize}
\indent\indent Next-generation telescopes coming online in
the 2020s will provide significant improvements in spectral resolution
and signal-to-noise, allowing for spectral studies of smaller
KBOs. Characterization of the surface compositions of small to
intermediate-sized KBOs will further enhance our understanding of how
the various physical and chemical processes shape the surfaces of objects
in the Kuiper Belt.}

\section*{Acknowledgements}
\onehalfspace{We appreciate the work of the anonymous
  reviewer whose helpful comments improved this paper. A very special
  thanks to everyone who was part of the
  schedule reorganization in July 2015, including S. Dahm, A. Howard,
  C. Jordan, H. Knutson, G. Marcy, and M. Rich. We would also like to
  thank the Support Astronomers and Observing Assistants at the
  W. M. Keck Observatory, without whom this work would not have been possible: R. Campbell,
S. Dahm, G. Doppman, H. Hershey, J. McIlroy, J. Rivera, and
H. Tran. Bobby Bus and Rick Binzel provided useful advice for ensuring
an effective telluric correction. The data presented herein were
obtained at the W.M. Keck Observatory, which is operated as a
scientific partnership among the California Institute of Technology,
the University of California, and the National Aeronautics and Space
Administration. The Observatory was made possible by the generous
financial support of the W.M. Keck Foundation. A portion of the data presented
herein were obtained using the UCI Remote Observing
Facility, made possible by a generous gift from John and Ruth Ann
Evans. We wish to recognize and acknowledge the significant cultural
role and reverence of the summit of Mauna Kea within the indigenous
Hawaiian community and to express our appreciation for the opportunity
to observe from this special mountain. This work was funded by NASA
PAST NNX13AG06G, NASA NESSF 14-PLANET14F-0045, and NASA NESSF 15-PLANET15R-0023.}

\section*{Appendix A}
\subsection*{A.1. Data reduction}
\onehalfspace{This appendix provides a more detailed description of
  the reduction process for the OSIRIS data; Fig. 10 is a flowchart
  that details the steps in the reduction. The raw OSIRIS data
  consisted of 1,019 overlapping spectra on the detector. The first step in the reduction process was to run
the data through the OSIRIS Data Reduction Pipeline (Krabbe et al.,
2004). The pipeline performs many tasks, including removing
cosmic rays, subtracting a bias offset,
and correcting dispersion. There is also an option for subtraction of
another image, such as a dark frame, prior to image rectification. The
primary purpose of the pipeline is to separate the overlapping spectra
using a process similar to a Lucy-Richardson deconvolution
(Richardson, 1972; Lucy, 1974). This is achieved using rectification
matrices, maps of the point spread function of each lenslet at every
wavelength, and each combination of filter and plate scale has a unique
rectification matrix. The final product of the pipeline is a 3D data
cube with two spatial and one spectral dimension. Each 2D
image in the data cube corresponds to a different wavelength. We
produced two data cubes from each raw file: one by subtracting a
master dark frame (the median of 5 individual dark frames) of
suitable integration time (dark-subtracted) and one by subtracting the
sky image of that set (sky-subtracted).\\
\indent Following the pipeline reduction, we further reduced the data
using an in-house IDL routine to extract 1D Pluto, Charon, and solar analog
spectra. All data cubes were reduced following the process outlined below. The
first step was to compute the mean of the sky-subtracted 3D data
cube in the spectral direction. This collapsed the data cube into a 2D
image (Fig. 2, left-hand side). In order to simplify the reduction process, we chose to use the
full width at half maximum (FWHM) and centroid positions of Pluto and Charon from the 2D
average image. The centroid
position was set as the position of the trace and the FWHM as the
aperture radius in every image of the data
cube. To ensure that this was acceptable, we examined the
centroid position of the solar analog in a few Hbb and Kbb data
cubes; the maximum deviation of the centroid position from the average
within any given cube was about 0.1 pixel. The aperture radii for Pluto, Charon, and the solar analog were
different; using the same aperture size for Pluto and Charon would
have introduced additional noise into the Charon spectrum. Any pixels
not at least partially included in the circular
apertures of Pluto or Charon were considered background
pixels. The median value of the background pixels in each image was
then subtracted from every pixel in the image. The Pluto and Charon fluxes
at each wavelength were extracted using the IDL routine
\textit{basphote}\renewcommand{\thefootnote}{\fnsymbol{footnote}}\footnote{The
  \textit{basphote} IDL routine was written by Marc Buie and performs
  circular aperture photometry to obtain a flux value for an object
  in a 2D image. Documentation for
  \textit{basphote} can be found at\\ \url{http://www.boulder.swri.edu/~buie/idl/pro/basphote.html}.},
resulting in a 1D spectrum. The same procedure was followed to extract
the 1D spectrum of the solar analog.

\begin{figure}[h!]
\begin{center}
\includegraphics[scale=0.75,trim=0cm 0.5cm 0cm 0.5cm,clip=true]{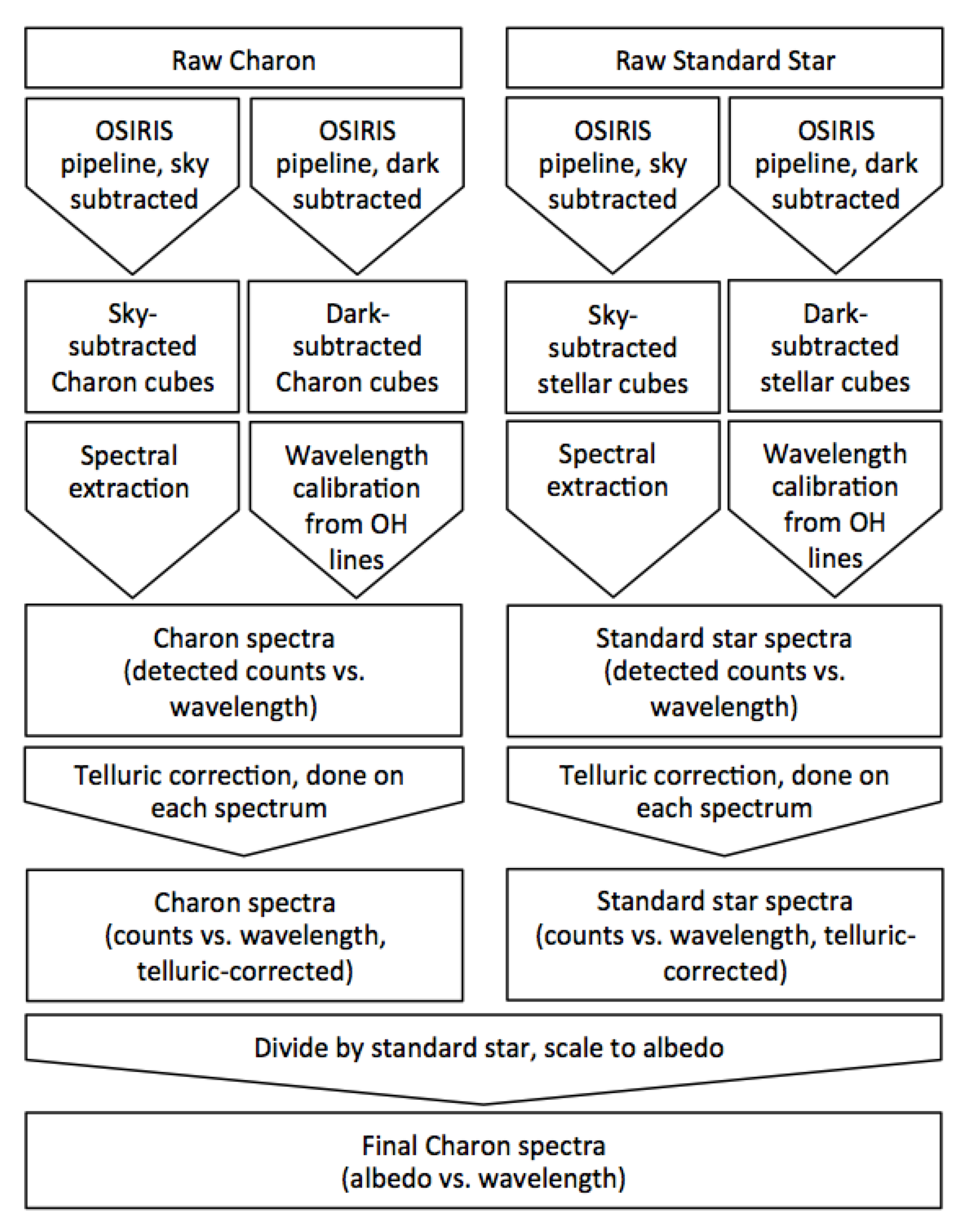}
\caption{Flowchart detailing the reduction strategy.}
\end{center}
\end{figure}

\indent We performed our own wavelength calibration using the
dark-subtracted sky frames from each set of data cubes. These cubes
were collapsed into a 1D spectrum by summing the pixels along both
spatial dimensions. A third order polynomial was robustly fit to the
1D spectrum to model the increase in sky glow with increasing
wavelength, especially in the $K$ band. The polynomial was then
subtracted from the spectrum, yielding a flat continuum between the
emission lines. The average and standard deviation of the continuum
were robustly determined (by removing outliers as in Buie and Bus,
1992) in order to prevent the emission lines from
skewing the values. To identify the OH emission lines, a threshold was
set at the average of the continuum plus 5 standard deviations. We
fit a Lorentzian profile to each group of points above the threshold
to determine the centroid of the peak. These values were matched to
tabulated vacuum wavelengths of OH emission from Rousselot et al. (2000) to
construct a wavelength solution. This procedure was identical for the
Pluto/Charon and solar analog data cubes. We used the wavelength solutions
from this procedure, as they were nearly
identical to the solutions provided by the Data Reduction Pipeline.

\subsection*{A.2. Telluric correction}
\indent Telluric absorption from water vapor in Earth's atmosphere was
corrected for in both $H$ and $K$ spectra using an iterative
method developed in-house and written in IDL. The basic process was to
determine the water vapor overburden that best corrected for telluric
absorption in the separate Hbb and Kbb solar analog spectra on a
nightly basis;
we then calculated the transmission spectra at the various airmasses of the individual
Charon spectra for that water vapor overburden. ATRAN models (Lord,
1992) were generated using a web-based input
form\renewcommand{\thefootnote}{\fnsymbol{footnote}}\footnote{\url{http://atran.sofia.usra.edu/cgi-bin/atran/atran.cgi}}.
These models are synthetic atmospheric transmission spectra that take into
account atmospheric absorption primarily from water vapor, but also
from trace species such as ozone and methane. The
parameters that go into creating the ATRAN models are observatory altitude,
observatory latitude, water vapor overburden, number of atmospheric
layers, zenith angle, wavelength range, and smoothing resolution. The
observatory altitude (13,600 feet for Keck Observatory), observatory
latitude (39$^{\circ}$, as recommended), number of atmospheric layers (2, as recommended),
wavelength range (1.473-2.382 $\mu$m), and smoothing resolution (3800)
were the same for all ATRAN models. Even though Keck is actually at a
latitude of 20$^{\circ}$, the instructions recommended setting the
latitude to 39$^{\circ}$. This value determines the quantity of
ozone included in the model and ozone dependence with latitude is
negligible compared to diurnal and seasonal variations. ATRAN models were
generated by varying the water vapor
overburden from 250 $\mu$m to 10,000 $\mu$m in steps of 250
$\mu$m for each value of the zenith angle. The zenith angles ($z$=cos$^{-1}$(1/$X$)) used were
those of the solar analog spectra and were
calculated from the specified airmass value ($X$)
provided in the data cube FITS headers. The ATRAN
atmospheric absorption models were then resampled onto the $H$ and $K$
wavelength grids. Our iterative method, \textit{btellcor}, takes an
uncorrected 1D spectrum as an input. From there, it determines the
wavelength range of the spectrum and its airmass, then selects the
appropriate set ($H$ or $K$) of resampled ATRAN
models. The uncorrected spectrum is divided by each model in the set.
A line is fit across a region of strong telluric absorption
(1.753-1.803 $\mu$m for $H$ and 2.315-2.377 $\mu$m for $K$), with the
model that best eliminates telluric absorption determined by
calculating the standard deviation of the residuals (data minus
model). Because the original
ATRAN models were created at a relatively coarse interval of water
vapor overburdens (250
$\mu$m), the true best fit model may lie in the gaps between ATRAN
models. To more accurately determine the best fit water vapor
overburden, new ATRAN models are constructed at 10 $\mu$m intervals between the ATRAN models
bracketing the minimum standard deviation model. The model that best eliminates telluric absorption
is then determined in the same manner as described previously.\\
\indent We assumed a minimal variation in water vapor overburden
throughout each 4-5 hour observing session. This assumption was backed
up by optical depth measurements from the Caltech Submillimeter
Observatory
(CSO)\renewcommand{\thefootnote}{\fnsymbol{footnote}}\footnote{\url{http://cso.caltech.edu/tau/}};
mm of H$_2$O was calculated by multiplying the optical depth obtained
at 225 GHz by 20. Of the 6 observing sessions, the water vapor
overburden changed by less than 1 mm for 4 sessions (2015-07-24,
2015-07-28, 2015-08-08, and 2015-08-30 UT), about 1 mm for
1 session (2015-07-18 UT), and 1
night did not have data available (2015-07-14 UT).\\
\indent Using \textit{btellcor},
the water vapor overburden that best eliminated telluric absorption
for each filter/night combination was determined from the average of
the 2 A and 1 B spectra of the solar analog obtained at the beginning of the
night; the airmass was taken from the B spectrum. We did not aim to
determine the true value of the water vapor overburden, just the value
that best corrected the spectrum. ATRAN models were generated using
the web-based interface at the determined water vapor overburden and
the zenith angle of each Charon spectrum. Each telluric-corrected
Charon spectrum was then divided by the appropriate corrected solar
analog spectrum to eliminate solar absorption bands.

\subsection*{A.3. Geometric albedo scaling}
\indent Conversion of relative reflectance to geometric albedo
required scaling to a previously published Charon spectrum due to
light losses from the Keck AO system. The maximum Strehl ratios for the $H$ and
$K$ bands with NGS are 0.45 and 0.65, respectively\renewcommand{\thefootnote}{\fnsymbol{footnote}}\footnote{\url{http://www2.keck.hawaii.edu/optics/ngsao}};
for LGS, the maximum Strehl ratio is
0.39\renewcommand{\thefootnote}{\fnsymbol{footnote}}\footnote{\url{http://www2.keck.hawaii.edu/optics/lgsao/performance.html}}. The average in the
relatively flat regions 1.72-1.80 $\mu$m ($H$ band) and 2.23-2.26
$\mu$m ($K$ band) were calculated in our grand average spectrum. We
used the HST/NICMOS Charon spectrum from Fig. 3 of Buie and Grundy
(2000) to anchor our geometric albedo calculation: between 1.72-1.80
$\mu$m and 2.23-2.26 $\mu$m the albedo values are 0.365 and 0.32, respectively. The $H$ spectrum
was then multiplied by 0.365 and divided by the $H$ average; the $K$ spectrum was
multiplied by 0.32 and divided by the $K$ average. Buie and Grundy (2000) calculated
the geometric albedo using a Charon radius of 593 km; Stern et
al. (2015) provide a more accurate value of 606 km from New Horizons
observations. The $H$ and $K$ spectra were both scaled by (593 km/606 km)$^2$ to
correct the geometric albedo for the difference in radii (albedo is
proportional to the inverse of the square of the radius).}

\section*{Appendix B. Supplementary material}
\onehalfspace{Supplementary data associated with this article can be
  found, in the online version, at INSERT URL HERE.}

\section*{References}

\noindent Barkume, K.M., Brown, M.E., Schaller, E.L., 2006. Water ice
on the satellite of Kuiper Belt Object 2003 EL$_{61}$. ApJ 640: L87-L89.\\
\\
Barucci, M.A., et al., 2008. Surface composition and
temperature of the TNO Orcus. A\& 479, L13-L16.\\
\\
Barucci, M.A., Dalle Ore, C.M., Perna, D., Cruikshank, D.P.,
Doressoundiram, A., Alvarez-Candal, A., Dotto, E., Nitschelm, C.,
2015. (50000) Quaoar: Surface composition variability. A\&A 584, A107.\\
\\
Brown, R.H., Cruikshank, D.P., Tokunaga, A.T., Smith, R.G., Clark,
R.N., 1988. Search for volatiles on icy satellites. I-Europa. Icarus
74, 262-271.\\
\\
Brown, M.E., Calvin, W.M., 2000. Evidence for crystalline
water and ammonia ices on Pluto's satellite
Charon. Science 287, 107-109.\\
\\
Brown, M.E., Barkume, K.M., Ragozzine, D., Schaller, E.L., 2007. A
collisional family of icy objects in the Kuiper belt. Nature 446, 294-296.\\
\\
Brozovi\'c, M., Showalter, M.R., Jacobson, R.A., Buie, M.W., 2015. The
orbits and masses of satellites of Pluto. Icarus 246, 317-329.\\
\\
Brucker, M.J., et al., 2009. High albedos of low inclination Classical
Kuiper belt objects. Icarus 201, 284-294.\\
\\
Buie, M.W., Cruikshank, D.P., Lebofsky, L.A., Tedesco, E.F.,
1987. Water frost on Charon. Nature 329, 522-523.\\
\\
Buie, M.W., Bus, S.J., 1992. Physical observations of (5145) Pholus. Icarus 100, 288-294.\\
\\
Buie, M.W., Grundy, W.M., 2000. The distribution and physical state
of H$_2$O on Charon. Icarus 148, 324-339.\\
\\
Buie, M.W., Grundy, W.M., Young, E.F., Young, L.A., Stern, S.A.,
2010. Pluto and Charon with the \textit{Hubble Space
  Telescope}. I. Monitoring global change and improved surface
properties from light curves. AJ 139, 1117-1127.\\
\\
Buratti, B.J., et al., 2016. Global albedos of Pluto and Charon from
LORRI \textit{New Horizons} observations. Icarus, submitted.\\
\\
Canup, R.M. 2011. On a giant impact origin of Charon, Nix, and Hydra. AJ 141, 35-44.\\
\\
Christy, J.W., Harrington, R.S., 1978. The satellite of Pluto. AJ 83, 1005-1008.\\
\\
Cook, J.C., Desch, S.J., Roush, T.L., Trujillo, C.A., Geballe, T.R.,
2007. Near-infrared spectroscopy of Charon: Possible evidence for
cryovolcanism on Kuiper Belt Objects. AJ 663, 1406-1419.\\
\\
Cook, J.C., et al., 2016. Spectroscopy of Pluto's small satellites. Joint AAS/Division for
Planetary Sciences 48/European Planetary Science Congress 11 Meeting
Abstracts, \#205.03.\\
\\
Cooper, J.F., Christian, E.R., Richardson, J.D., Wang, C.,
2003. Proton irradiation of centaur, Kuiper Belt, and Oort Cloud
objects at plasma to cosmic ray energy. Earth Moon and Planets 92, 261-277.\\
\\
Dalle Ore, C.M., et al., 2016. Charon's near IR ice signature as seen
by New Horizons. 47$^{th}$ LPSC, No. 1903, p. 2122.\\
\\
de Bergh, C., Delsanti, A., Tozzi, G.P., Dotto, E., Doressoundiram,
A., Barucci, M.A., 2005. The surface of the transneptunian object
90482 Orcus. A\&A 437, 1115-1120.\\
\\
DeMeo, F.E., Dumas, C., Cook, J.C., Carry, B., Merlin, F., Verbiscer,
A.J., Binzel, R.P., 2015. Spectral variability of Charon's 2.21-$\mu$m
feature. Icarus 246, 213-219.\\
\\
Dumas, C., Terrile, R.J., Brown, R.H., Schneider, G., Smith, B.A.,
2001. Hubble Space Telescope NICMOS spectroscopy of Charon's leading
and trailing hemispheres. AJ 121, 1163-1170.\\
\\
Dumas, C., Carry, B., Hestroffer, D., Merlin, F., 2011. High-contrast
observations of (136108) Haumea. A crystalline water-ice multiple
system. A\&A 528, 1-6.\\
\\
Fornasier, S., Dotto, E., Barucci, M.A., Barbieri, C., 2004. Water ice
on the surface of the large TNO 2004 DW. A\&A 422, L43-L46.\\
\\
Fraser, W.C., et al., 2013. Limits on Quaoar's atmosphere. ApJL 774, L18-L21.\\
\\
Fray, N., Schmitt, B., 2009. Sublimation of ices of astrophysical
interest: A bibliographic review. P\&SS 57, 2053-2080.\\
\\
Grundy, W.M., Schmitt, B., 1998. The temperature-dependent
near-infrared absorption spectrum of hexagonal H$_2$O ice. JGR 103, 25809-25822.\\
\\
Grundy, W.M., et al., 2016. Surface compositions across Pluto and
Charon. Science 351, aad9189.\\
\\
Houk, N., Smith-Moore, M., 1988. Michigan Catalogue of Two-dimensional
Spectral Types for the HD Stars. Volume 4, Declinations
-26$^{\circ}$.0 to -12$^{\circ}$.0. Department of Astronomy, University of Michigan, Ann Arbor, MI.\\
\\
Jewitt, D.C., Luu, J., 2004. Crystalline water ice on the Kuiper Belt
object (50000) Quaoar. Nature 432, 731-733.\\
\\
Johnson, R.E., Oza, A., Young, L.A., Volkov, A.N., Schmidt, C.,
2015. Volatile loss and classification of Kuiper Belt Objects. ApJ
809, 43-51.\\
\\
Krabbe, A., et al., 2004. Data reduction pipeline for OSIRIS, the new
NIR diffraction-limited imaging field spectrograph for the Keck
adaptive optics system. SPIE 5492, 1403-1410.\\
\\
Larkin, J., et al., 2006. OSIRIS: A diffraction limited integral field
spectrograph for Keck. Proc. SPIE 6269, 42-46.\\
\\
Lellouch, E., Stansberry, J., Emery, J., Grundy, W., Cruikshank, D.P.,
2011. Thermal properties of Pluto's and Charon's surfaces from
\textit{Spitzer} observations. Icarus 214, 701-716.\\
\\
Leto, G., Baratta, G.A., 2003. Ly-$\alpha$ photon induced
amorphization of Ic water ice at 16 Kelvin. A\&A 397, 7-13.\\
\\
Livingston, F.E., Smith, J.A., George, S.M., 2002. General trends for
bulk diffusion in ice and surface diffusion on ice. J. Phys. Chem. A
106, 6309-6318.\\
\\
Lord, S.D., 1992. A new software tool for computing Earth's
atmospheric transmission of near- and far-infrared radiation. NASA
Technical Memorandum 103957.\\
\\
Lucy, L.B., 1974. An iterative technique for the rectification of
observed distributions. AJ 79, 745-754.\\
\\
Marcialis, R.L., Rieke, G.H., Lebofsky, L.A., 1987. The surface
composition of Charon--Tentative identification of water ice. Science
237, 1349-1351.\\
\\
Mastrapa, R.M.E., Brown, R.H., 2006. Ion irradiation of crystalline
H$_2$O-ice: Effect on the 1.65-$\mu$m band. Icarus 183, 207-214.\\
\\
Mastrapa, R.M., Bernstein, M.P., Sandford, S.A., Roush, T.L.,
Cruikshank, D.P., Dalle Ore, C.M., 2008. Optical constants of
amorphous and crystalline H$_2$O-ice in the near infrared from 1.1 to
2.6 $\mu$m. Icarus 197, 307-320.\\
\\
Merlin, F., Guilbert, A., Dumas, C., Barucci, M.A., de Bergh, C.,
Vernazza, P., 2007. Properties of the icy surface of the TNO 136108
(2003 EL$_{61}$). A\&A 466, 1185-1188.\\
\\
Merlin, F., Barucci, M.A., de Bergh, C., DeMeo, F.E., Alvarez-Candal,
A., Dumas, C., Cruikshank, D.P., 2010. Chemical and physical
properties of the variegated Pluto and Charon surfaces. Icarus 210, 930-943.\\
\\
Mieda, E., et al., 2014. Efficiency measurements and installation of a
new grating for the OSIRIS spectrograph at Keck Observatory. PASP 126,
250-263.\\
\\
Moore, J.M., et al., 2016. The geology of Pluto and Charon through the
eyes of New Horizons. Science 351, 1284-1293.\\
\\
Moore, M.H., Ferrante, R.F., Hudson, R.L., Stone, J.N., 2007. Ammonia
water ice laboratory studies relevant to outer Solar System
surfaces. Icarus 190, 260-273.\\
\\
Pinilla-Alonso, N., Brunetto, R., Licandro, J., Gil-Hutton, R., Roush,
T.L., Strazzulla, G., 2009. The surface of (136108) Haumea (2003
EL$_{61}$), the largest carbon-depleted object in the trans-Neptunian
belt. A\&A 496, 547-556.\\
\\
Press, W.H., Teukolsky, S.A., Vetterling, W.T., Flannery, B.P.,
2007. Numerical Recipes: The Art of Scientific Computing, third
ed. Cambridge University Press.\\
\\
Rabinowitz, D.L., Schaefer, B.E., Tourtellotte, S.W., 2007. The
diverse solar phase curves of distant icy bodies. I. Photometric
observations of 18 trans-Neptunian objects, 7 Centaurs, and Nereid. AJ
133, 26-43.\\
\\
Reuter, D.C., et al., 2008. Ralph: A visible/infrared imager for the
New Horizons Pluto/Kuiper Belt mission. Space Sci. Rev. 140, 129-154.\\
\\
Richardson, W.H., 1972. Bayesian-based iterative method of image
restoration. JOSA 62, 55-59.\\
\\
Rousselot, P., Lidman, C., Cuby, J.-G., Moreels, G., Monnet, G.,
2000. Night-sky spectral atlas of OH emission lines in the
near-infrared. A\&A 354, 1134-1150.\\
\\
Schaller, E.L., Brown, M.E., 2007. Volatile loss and retention on
Kuiper Belt Objects. ApJ 659, L61-L63.\\
\\
Sicardy, B., et al., 2011. A Pluto-like radius and a high albedo for
the dwarf planet Eris from an occultation. Nature 478, 493-496.\\
\\
Stern, S.A., et al., 2015. The Pluto system: Initial results from its
exploration by New Horizons. Science 350, 292-299.\\
\\
Stern, S.A., et al., 2016. New Horizons constraints on Charon's
present day atmosphere. arXiv:1608.06955.\\
\\
Strazzulla, G., Palumbo, M.E., 1998. Evolution of icy surfaces: An
experimental approach. P\&SS 46, 1339-1348.\\
\\
Tholen, D.J., Buie, M.W., Grundy, W.M., Elliott, G.T., 2008. Masses of
Nix and Hydra. AJ 135, 777-784.\\
\\
Trujillo, C.A., Brown, M.E., Barkume, K.M., Schaller, E.L.,
Rabinowitz, D.L., 2007. The surface of 2003 EL$_{61}$ in the
near-infrared. AJ 655, 1172, 1178.\\
\\
Uras, N., Devlin, J.P., 2000. Rate study of ice particle conversion to
ammonia hemihydrate: Hydrate crust nucleation and NH$_3$
diffusion. J. Phys. Chem. A 104, 5770-5777.\\
\\
Verbiscer, A.J., Peterson, D.E., Skrutskie, M.F., Cushing, M.,
Helfenstein, P., Nelson, M.J., Smith, J.D., Wilson, J.C.,
2006. Near-infrared spectra of the leading and trailing hemispheres of
Enceladus. Icarus 182, 211-223.\\
\\
Verbiscer, A.J., Peterson, D.E., Skrutskie, M.F., Cushing, M., Nelson,
M.J., Smith, J.D., Wilson, J.C., 2007. Simultaneous spatially-resolved
near-infrared spectra of Pluto and Charon. LPSC 38, 2318.\\
\\
Young, L.A., et al., 2008. New Horizons: Anticipated scientific
investigations at the Pluto system. Space Sci. Rev. 140, 93-127.\\
\\
Zangari, A., 2015. A meta-analysis of coordinate systems and
bibliography of their use on Pluto from Charon's discovery to the
present day. Icarus 246, 93-145.\\
\\
Zheng, W., Jewitt, D., Kaiser, R.I., 2009. On the state of water ice
on Saturn's moon Titan and implications to icy bodies in the outer
Solar System. J. Phys. Chem. A 113, 11174-11181.

\end{document}